# Abnormally large changes in anharmonicity of intramolecular vibrations in free clusters of nitrogen

**O. Danylchenko, Yu. Doronin, S. Kovalenko,**
**M. Libin, V. Samovarov, V. Vakula**[*]

*B. Verkin Institute for Low Temperature Physics and Engineering*
*of the National Academy of Sciences of Ukraine,*
*47 Lenin Ave., Kharkiv, 61103, Ukraine*

**Introduction**

Energy characteristics of a molecule placed in a matrix are known to differ from those in the gaseous phase. The matrix shift is widely studied for various molecules (included those of atmospheric gases) imbedded in bulk rare gas solids [1]. However for small clusters, the effect of matrix shift has been studied to a much lesser extent. Here we report our results on the change of vibrational anharmonicity and frequencies of nitrogen molecules in small free clusters of nitrogen.

**Experimental**

We made cathodoluminescence measurements on free clusters of nitrogen produced in the process of an isentropic expansion of a gas mixture flowing at a supersonic velocity into vacuum through a conical nozzle. The nozzle parameters were as follows: critical cross-section diameter – 0.34 mm, cone expansion angle – 8.6°, outlet-to-critical cross-section area ratio – 36.7. The gas jet was evacuated by a condensation pump cooled with liquid hydrogen. The pressure $p_0$ and temperature $T_0$ of the gas at the entrance of the nozzle were 1 atm and 170-210 K, respectively. The clusters were excited by an electronic beam of the energy of 1 keV, the cathodoluminescence spectra were registered in the region of 45,000-70,000 cm$^{-1}$ (5.5-8.7 eV). According to the data of electron diffraction study, the cluster temperature was about 40 K, the size did not exceed 1000 atoms. The clusters characterized by such parameters have a quasi-crystalline icosahedral structure with a 5-fold symmetry axis.

---

[*] Corresponding author: vakula@ilt.kharkov.ua

**Results**

The measured luminescence spectra (Fig. 1) of pure nitrogen clusters were dominated by the spectral series corresponding to the transitions w $^1\Delta_u \rightarrow$ X $^1\Sigma_g^+$ from an excited to the ground state of nitrogen molecules. In bulk nitrogen samples, these series are much less prominent in comparison with the intense transitions from the lower-lying term: a′ $\Sigma_u^- \rightarrow$ X $^1\Sigma_g^+$ [2]. We observed transitions $E_{v',v''}$ from the vibrational levels v′ = 0, 1, 2 of the upper term w $^1\Delta_u$ to the levels v″ = 1, 2…13 of the ground state X $^1\Sigma_g^+$ (see Fig. 2). A strong dependence of the matrix shift of the energy positions was found for the transitions $(E_{v',v''})^{cluster} - (E_{v',v''})^{gas}$ on v′ and v″, where $E_v = \hbar\omega_e(v + 1/2) - \hbar\omega_e x_e(v + 1/2)^2$ and $E_{v',v''} = E_{v'} - E_{v''}$. In bulk nitrogen samples, the shift is constant (see. Fig. 3). Analysis of the non-linear dependence of matrix shift showed that in clusters the anharmonicity parameters undergo unusually strong changes for both the upper and lower states compared to the bulk and gaseous samples. In gas, the vibration quantum for the X $^1\Sigma_g^+$ state is $\omega_e$ = 2359 cm$^{-1}$, while the anharmonicity parameter is $\omega_e x_e$ = 14.3 cm$^{-1}$, but for the clusters the relevant values inferred from the spectra are $\omega_e$ = 2356 cm$^{-1}$ and $\omega_e x_e$ = 7.2 cm$^{-1}$, respectively; for the w $^1\Delta_u$ state the parameters are $\omega_e$ = 1559 cm$^{-1}$ and $\omega_e x_e$ = 11.6 cm$^{-1}$ (gas), and $\omega_e$ = 1799 cm$^{-1}$ and $\omega_e x_e$ = 31.6 cm$^{-1}$ (clusters). Such strong changes (several times in magnitude) in anharmonicity parameter upon a transition from gas to solid have not been observed so far. Earlier, the strongest changes of vibration parameters that we know were reported for the case of absorption to the excited state B $^3\Sigma_u^-$ in oxygen molecules embedded a krypton matrix (increase in $\omega_e x_e$ up to ≈50% for the excited state with practically unchanged value for the ground state X $^3\Sigma_g^-$) [3].

**Conclusion**

For the first time the influence of cluster environment on frequency and anharmonicity of the vibration of nitrogen molecules has been studied experimentally. It is shown that in the small clusters (having the radius of ≈15 Å and consisting of ≈1000 atoms) there is an abnormally large (several times in magnitude) change of anharmonicity (absent in bulk samples), while the vibrational frequencies remain almost unchanged. Analysis of the results obtained has demonstrated that icosahedral clusters, in contrast to bulk samples of cubic symmetry, can be characterized by a strong internal potential responsible for the abnormal changes in the vibration parameters of nitrogen molecules.

**Figure captions**

**Fig. 1.** A typical cathodoluminescence spectrum from pure nitrogen clusters observed in our experiments. The arrows indicate three series of transitions from the levels $v' = 0, 1, 2$ of the excited state w $^1\Delta_u$ to the $v''$-th level of the ground state X $^1\Sigma_g^+$ (the number $v''$ is shown near the relevant arrow). The two strongest lines in the spectrum are due to atomic luminescence.

**Fig. 2.** A schematic picture of the transitions from the excited state w $^1\Delta_u$ to the ground state X $^1\Sigma_g^+$ in nitrogen clusters. The experimentally detected transitions of the three series w $^1\Delta_u$ (v′) → X $^1\Sigma_g^+$ (v″) for $v' = 0, 1, 2$ and $v'' = 1, 2… 13$ are shown.

**Fig. 3.** The matrix shift $E_{cl}$-$E_{gas}$ of the three types of transitions w $^1\Delta_u$ (v′) → X $^1\Sigma_g^+$ (v″) as a function of number of the final state in clusters (our measurements, filled symbols) and in bulk solid samples (taken from Ref. [4], open symbols). Solid lines are fitting curves to the matrix shift calculated on the basis of modificated anharmonicity parameter and vibrational frequency, see the text.

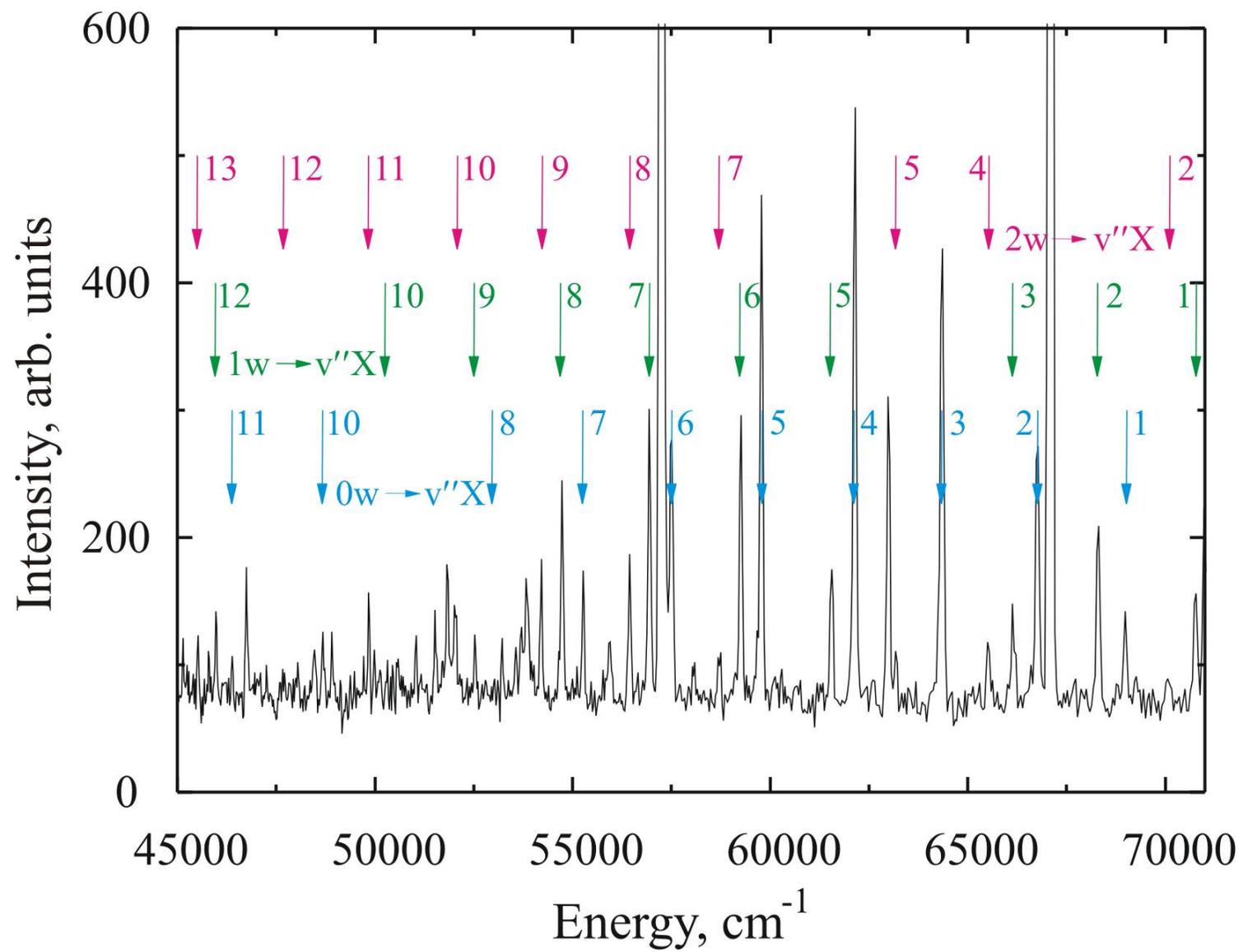

**Fig. 1**

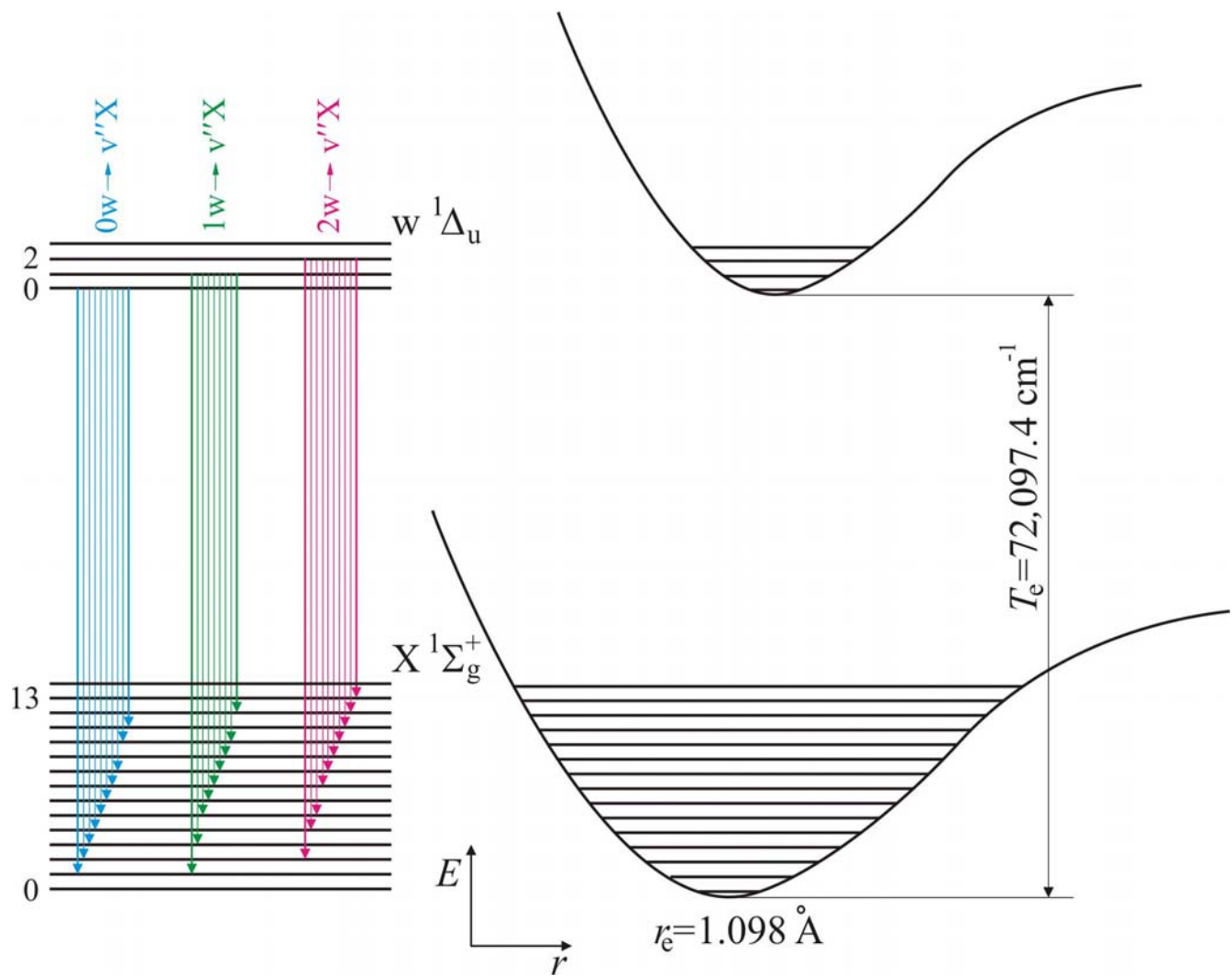

**Fig. 2**

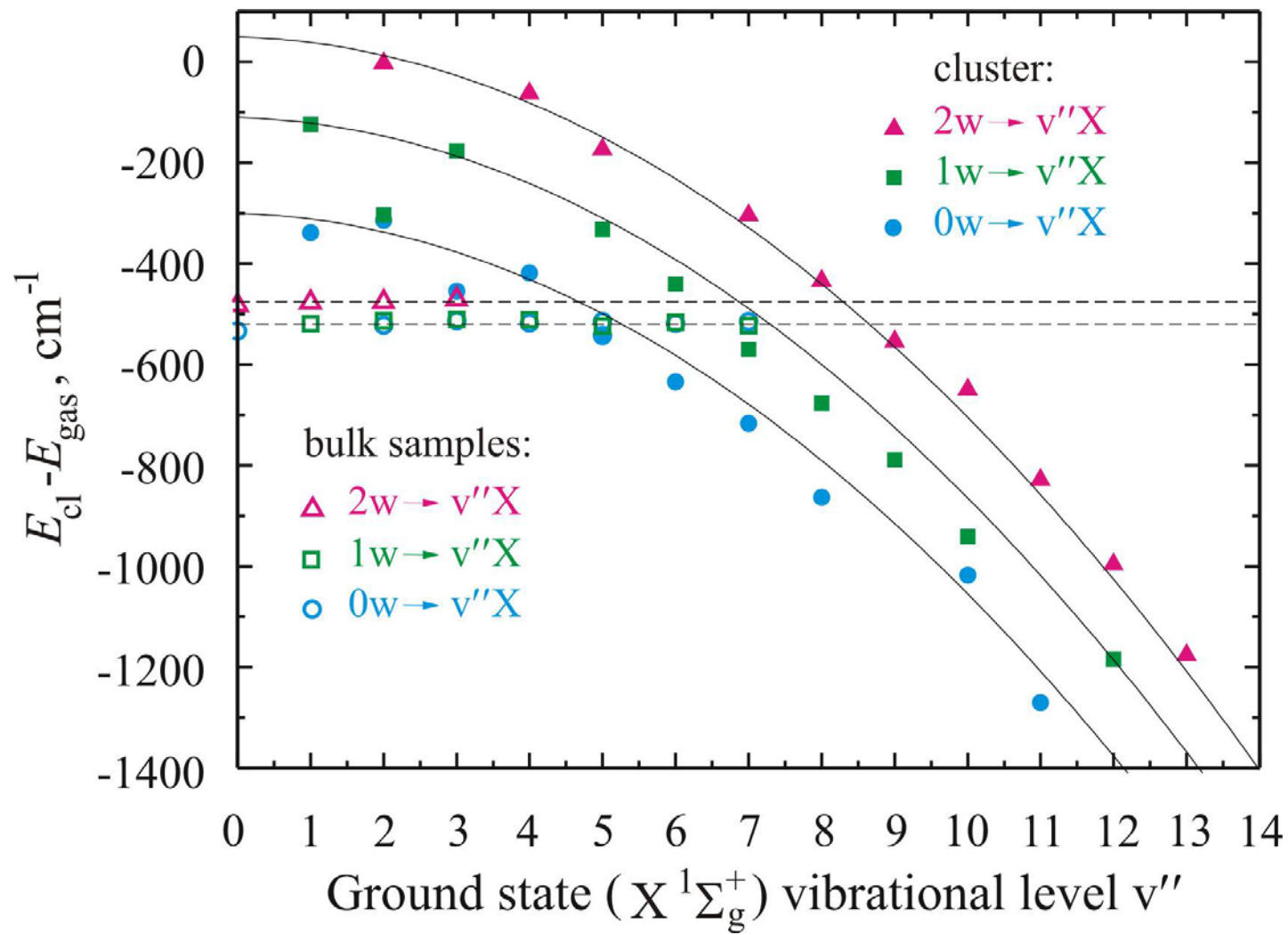

**Fig. 3**